\documentstyle[aps,prl,epsfig,floats]{revtex}

\def\gsim{\lower0.5ex\hbox{$\:\buildrel >\over\sim\:$}}
\def\lsim{\lower0.5ex\hbox{$\:\buildrel <\over\sim\:$}}
\def \rp{{R\hspace{-0.22cm}/}_P}
\def \lp{{L\!\!\!/}}
\def \n{\noindent}

\begin{document}

\title{Sneutrino--Higgs mixing in \boldmath{$WW$} and 
\boldmath{$ZZ$} production in supersymmetry with R-parity 
violation}
\author{S. Bar-Shalom$^a$, G. Eilam$^b$ and B. Mele$^a$}
\address{$^a$ INFN, Sezione di Roma and Dept. of Physics, 
University of Roma I, La Sapienza, Roma, 
Italy.\\
$^b$ Physics Department, Technion-Institute of Technology, Haifa 32000, 
Israel.}
\date{\today }
\maketitle

\begin{abstract}
We consider new $s$-channel scalar exchanges in 
$e^+e^- \to ZZ,~W^+W^-$ in supersymmetry with a small 
lepton number violation. 
We show that a small bilinear $R$-parity violating term  
which leads to sneutrino--Higgs mixing can give rise to a 
significant scalar resonance enhancement in $e^+e^- \to ZZ,~W^+W^-$. 
We use the LEP2 measurements of the $WW$ and $ZZ$ 
cross-sections to place useful constraints on this scenario. 
We also find, under conservative assumptions on the relevant parameter 
space involved, that
such an exchange of the sneutrino-like admixture in 
$e^+e^- \to ZZ,~W^+W^-$ 
may be accessible to a 500 GeV $e^+e^-$ collider.
\end{abstract}

\draft
\pacs{PACS numbers: 12.60.Jv, 11.30.Fs, 14.80.Cp, 13.85.Lg}


In spite of the indisputable
 success of the Standard Model (SM) in confronting 
experimental data, there are strong theoretical motivations 
for new beyond the SM physics being just around the corner. 
One of the most attractive new physics scenarios is supersymmetry (SUSY),  
which offers a plethora of new phenomena that might be observed in
upcoming future colliders. 
As opposed to the SM, lepton number does not have to be conserved 
in the SUSY Lagrangian. In fact, there is no fundamental principle that 
enforces lepton number conservation.

The SUSY superpotential can violate lepton number 
(or more generally R-parity)
via 
an R-parity violating (RPV) Yukawa-like trilinear term 
(RPVTT) in the purely leptonic sector, 
and via a mass-like RPV bilinear term (RPVBT) as follows 
\cite{rpreview}:\footnote{The RPVTT 
$\lambda^\prime \hat L \hat Q \hat {\bar D}$ is not 
relevant for our discussion.}

\begin{equation}
W_{\rp,\lp} \supset \epsilon_{ab} \left[ 
\lambda_{ijk} 
{\hat L}^a_i {\hat L}^b_j {\hat {\bar E}}_k/2 - \mu_i {\hat L}^a_i 
\hat H_u^b \right] ~,
\label{lp}     
\end{equation}

\n where $i,j,k =1,2$ or $3$ label the lepton generation and $a,b=1,2$ 
are SU(2) indices. 
In what follows we will 
assume that only $\mu_3 \ne 0$.\footnote{The effects of $\mu_1 \ne 0$ 
and/or $\mu_2 \ne 0$ are not crucial for the main outcome of this paper.}
The scalar 
potential contains the corresponding soft SUSY breaking RPV terms
in addition to the usual R-parity conserving (RPC) ones. 
The relevant ones for our discussion 
are $b_3 \tilde L_3 H_u$ and 
$M^2_{\tilde L H} \tilde L_3 H_d$ \cite{GH,davidson} which lead to 
a non-vanishing VEV of the tau-sneutrino, $v_3$.
However, since lepton number is not a conserved 
quantum number in this scenario, the $\hat H_d$ and $\hat L_3$ 
superfields loose
their identity and 
can be rotated to a particular basis ($\hat H_d^\prime,\hat L_3^\prime$) 
in which either $\mu_3$ or $v_3$ is zero \cite{basis}. 
In what follows, we find it convenient to choose the basis where $v_3=0$. 

Throughout this paper we will
assume a small lepton number violation in the 
SUSY Lagrangian. 
That is, $|\mu_3/\mu_0| \le 0.1$, $|\lambda_{ijk}| \le 0.1$ 
and $b_3/b_0 \le 0.1$, where $\mu_0$ is the usual RPC 
Higgs mass term, $\mu_0 \hat H_d \hat H_u$, 
and $b_0 {\tilde H}_d {\tilde H}_u$ is the corresponding soft term.
Note that in the $v_3=0$ basis
the minimization of the scalar potential 
yields \cite{GH}: 
$b_3=(M^2_{\tilde L H} + \mu_3 \mu_0) \cot\beta$, where 
$\tan\beta \equiv v_u/v_d$. Thus, 
in the general case, $b_3$ needs not 
vanish even if 
$\mu_3$ is vanishingly small, 
as may be suggested by low energy  
flavor changing processes (see e.g., \cite{hepph0005262}) and/or 
flavor changing $Z$-decays (see e.g., Bisset 
{\it et al.}, in \cite{neutrinomass}).
That is, if $M_{\tilde L H}^2 \gg \mu_3 \mu_0$ due to 
$\mu_3 \to 0$, then $b_3 \sim M_{\tilde L H}^2 \cot\beta$, in which case
RPV in the scalar sector decouples 
from the RPV in
the superpotential (i.e., $\mu_3$). 
Thus,   
small lepton number violation in the scalar 
potential 
can be realized by requiring only that 
$b_3 \ll b_0$.\footnote{Note that the laboratory limit on the $\tau$-neutrino
mass allows $b_3/b_0 \sim {\cal O}(1)$ \cite{davidson}}.

Some of the interesting phenomenological 
implications of the RPVBT are 
tree-level neutrino masses \cite{GH,basis,neutrinomass} and 
new scalar decay channels \cite{davidson,rpvbtdec}.
In this letter we suggest yet a new signature that can serve as an exclusive
probe of the RPVBT. 
In particular, one can have scalar resonances in massive 
gauge-boson pair production:

\begin{equation}
e^+e^- \to \Phi_k \to VV ~,~{\rm with}~~ V=W ~{\rm or}~ Z \label{eevv}~,
\end{equation}

\n where $\Phi_k$ are admixtures of the RPC CP-even neutral Higgs 
and tau-sneutrino fields 
as described below. Note that 
the CP-odd scalar states do not couple to $VV$ at tree-level. 
Such a resonance can arise with measurable consequences when 
the incoming $e^+e^-$ beam couples to the sneutrino component in 
$\Phi_k$ with a coupling $\propto \lambda \gg m_e/M_W$ in (\ref{lp}),  
while the $VV$ final state couples to the Higgs components in $\Phi_k$.
Therefore, this scalar exchange
can be attributed only to the 
Higgs--sneutrino mixing phenomena via the RPVBT and is a viable mechanism 
for probing a RPVBT beyond a RPVTT, i.e., 
$\sigma(e^+e^- \to \Phi_k \to VV) \to 0$ as $b_3 \to 0$. 
It should be  stressed that this resonance formation 
is different from previously suggested sneutrino resonances 
within RPV SUSY such as fermion pair 
production in leptonic 
colliders \cite{previousrpv}, 
since it is driven by RPV parameters 
in the soft breaking scalar sector and not purely by Yukawa-like RPV couplings 
in the superpotential.  

Let us define the SU(2) components of the neutral Higgs and stau 
fields, respectively, as:
$H^0_{d,u} \equiv (\xi_{d,u}^0 + v_{d,u} + i \phi_{d,u}^0)/\sqrt{2}$ and 
${\tilde\nu_\tau} \equiv (\tilde\nu_+^0 + v_3 + 
i \tilde\nu_-^0)/\sqrt{2}$, then setting $v_3=0$.
The CP-even $3\times 3$ symmetric 
scalar mass matrix is then obtained 
through the quadratic part of the scalar potential as: 
$\frac{1}{2} \Phi^0 M_+^2 (\Phi^0)^T$, where 
$\Phi^0 =( \xi_d^0,\xi_u^0,\tilde\nu_+^0)$.    

In the RPC limit  
the Higgs and sneutrino sectors decouple, i.e.,  
$M_+^2$ consists of 
the usual $2\times 2$ upper left block corresponding to the two 
CP-even Higgs states (see e.g., \cite{gunion}) and 
$(M_+^2)_{33} = m_{\tilde\nu_+^0}^2$, $(M_+^2)_{13,23}=0$. 
However, with $b_3 \ne 0$ and in the $v_3=0$ basis, $M_+^2$ acquires 
the new off-diagonal entries \cite{GH}: 
$(M_+^2)_{13}= (M_+^2)_{31}=b_3 \tan\beta$ and 
$(M_+^2)_{23}= (M_+^2)_{32}=-b_3$, 
which are 
responsible for the mixing 
of $\xi^0_{d,u}$ with $\tilde\nu_+^0$. 
As a result, the 
usual CP-even RPC Higgs states $H^0$ and $h^0$ 
($m_{H^0} > m_{h^0}$) 
acquire a small $\tilde\nu_+^0$
component and vice versa.\footnote{We use the superscript $0$ 
to denote the particle states in the RPC limit.}

The new CP-even scalar mass-eigenstates (i.e., the physical states) 
will be denoted here after by 
$\Phi \equiv (H,h,\tilde\nu_+)$, where, for small RPV in the SUSY 
Lagrangian, $H,h$ and $\tilde\nu_+$ are 
the states dominated by $H^0,h^0$ and $\tilde\nu_+^0$, respectively.    
They are related to the weak eigenstates 
via 
$\Phi^0_\ell = S_{\ell k} \Phi_k$, where $S$ is the rotation matrix 
that diagonalizes $M_+^2$, i.e.,  
$S^T M_+^2 S= {\rm diag}(m_H^2,m_h^2,m_{s \nu}^2)$ 
(throughout the rest of the paper we use $m_H,m_h,m_{s \nu}$ and 
$m_H^0,m_h^0,m_{s \nu}^0$
to denote the masses of the $H,h,\tilde\nu_+$ physical states and of 
the $H^0,h^0,\tilde\nu_+^0$ states, respectively).
The interaction vertices of the physical states $\Phi$ 
are then obtained by rotating the Feynman rules  
of the RPC SUSY Lagrangian (see e.g., \cite{rosiek}) 
with the matrix $S$. 
Thus, if $\Lambda_{\Phi^0_\ell}$ is an interaction 
vertex involving 
a weak state, then $\Lambda_{\Phi_k}$, the 
vertex involving the physical state, is given by
$\Lambda_{\Phi_k} = S_{\ell k} \Lambda_{\Phi^0_\ell}$.

Hence, the $\Phi_k VV$ coupling is given by:

\begin{equation}
\Lambda_{\Phi_k V_\mu V_\nu} = i (e/s_W) C_V m_V \left(c_\beta S_{1k} 
+s_\beta S_{2k} \right) g_{\mu \nu}~,
\end{equation}
    
\n where $C_V=1(1/c_W)$ for $V=W(Z)$, 
$s_W,c_W = \sin\theta_W,\cos\theta_W$ 
and $c_\beta,s_\beta=\cos\beta,\sin\beta$.
Note that for $b_3 \to 0$, 
$S_{11}=S_{22} \to \cos\alpha$,
$S_{12}=-S_{21} \to -\sin\alpha$, $S_{33} \to 1$ and $S_{13,23,31,32} \to 0$,
where $\alpha$ is the usual mixing angle of the RPC 
neutral CP-even Higgs sector \cite{gunion,djuadi}.

The couplings $\Lambda_{\Phi_k e^+e^-}$ are obtained 
from the RPVTT term in (\ref{lp}) and are 
$\Lambda_{\Phi_k e^+e^-} = S_{3k} 
\Lambda_{\tilde\nu_+^0 e^+e^-} = i S_{3k} \lambda_{131}/\sqrt 2$.
In our numerical analysis we will set 
$\lambda_{131}=0.1$ irrespective of $m_{s \nu}$.  
We note, though, that the present $1 \sigma$ limit \cite{rpreview},
$\lambda_{131} \lsim 0.06 \times m_{{\tilde e}_R}/[100~{\rm GeV}]$,  
does not rule out $\lambda_{131} \sim 0.3$ 
if the typical slepton mass is $m_{{\tilde e}_R} 
\sim m_{s \nu} \sim 500$ GeV; since 
$\sigma(e^+e^- \to \Phi \to VV) \propto \lambda_{131}^2$ (see below),  
it can be easily rescaled for different values of $\lambda_{131}$. 

$\sigma^0_V \equiv \sigma(e^+e^- \to \Phi \to VV)$ is thus given by:

\begin{eqnarray}
\sigma^0_V=\delta_V C_V^2 \frac{\alpha}{128 s_W^2}  
\frac{\beta_V (3-2 \beta_V^2 +3 \beta_V^4)}{s (1-\beta_V^2)}
\lambda_{131}^2 \times
\sum_{i,j=1}^3 S_{3i}S_{3j}A_iA_j \hat\Pi_i \hat\Pi_j^\star 
\label{sigma}~,
\end{eqnarray}

\n where $\delta_V=2(1)$ for $V=W(Z)$, $\beta_V = \sqrt{1-4M_V^2/s}$ 
($s$ is the square of the c.m. energy) and $A_k=\left(c_\beta S_{1k} 
+s_\beta S_{2k} \right)$. Also, $\hat\Pi_k=(1-x_k^2+ix_ky_k)^{-1}$, 
with $x_k = m_{\Phi_k}/\sqrt s$, $y_k = \Gamma_{\Phi_k}/\sqrt s$ 
and $\Gamma_{\Phi_k}$ is the $\Phi_k$ width. 
The interferences between the SM diagrams and our $s$-channel 
scalar diagrams are $\propto m_e$ and
therefore negligible. Thus, the total cross-section 
for $e^+e^- \to VV$ is simply the sum 
$\sigma^T_V=\sigma^{SM}_V + \sigma^0_V$. 

Let us now establish our relevant low-energy 
SUSY parameter space. 
The usual RPC CP-even 
Higgs sector can be described at tree-level by only 
two parameters \cite{gunion,djuadi}, conventionally chosen to be $m_A^0$ - the 
pseudo-scalar Higgs mass in the RPC case - and 
$t_\beta \equiv \tan\beta$. 
Furthermore, with the assumption of small RPV, i.e., 
${\rm RPV/RPC} \ll 1$, $m_A^0$ typically scales
as $(m_A^0)^2 \sim b_0 t_\beta$ for $t_\beta^2 \gg 1$.
Thus, without loss of generality we set $b_3 \equiv \varepsilon 
(m_A^0)^2 \cot\beta$,
such that small lepton number violation  
in the scalar sector is parameterized
by the dimensionless quantity $\varepsilon \sim b_3/b_0$.
Then $\varepsilon \ll 1$ corresponds to $b_3 \ll b_0$.
The parameter set $\left\{m_A^0,m_{s \nu}^0,t_\beta,\varepsilon \right\}$
therefore completely fixes $M_+^2$ at tree-level 
from which the rotation matrix $S$ and the tree-level masses $m_{\Phi_k}$ 
are derived.

We note that since $b_3= \varepsilon (m_A^0)^2 \cot\beta$, when 
$\varepsilon \ll 1$ [implying $b_3 \ll b_0$ and also $b_3 \ll (m_A^0)^2$],
the masses of the physical CP-even states $m_H,m_h,m_{s \nu}$ and of 
the CP-odd states (e.g., $m_A$) are only slightly shifted from 
the corresponding states in the RPC limit ($m_H^0,m_h^0,m_{s \nu}^0$
and $m_A^0$) as long as there is no accidental mass degeneracy among 
the scalar states \cite{futurepaper}. In particular, 
the scalar masses will be shifted 
by terms proportional to 
$b_3^2/[(m_{\Phi_k^0})^2 - (m_{\Phi_\ell^0})^2]$ with $k \neq \ell$ 
(see e.g., \cite{GH}). Thus, although we are using 
``bare'' masses (i.e., the scalar masses in the RPC 
limit) as inputs, it should be kept in mind that the physical masses are 
only slightly shifted. For example, for $\varepsilon =0.1$
and $|m_{s \nu}^0 - m_A^0|,~ |m_h^0 - m_A^0| \sim 100$ GeV we find that 
the shift in $m_A$, $|m_A - m_A^0|$, is at the level of a few 
percent at the most for both a low or a high $\tan\beta$ scenario (for 
more details see \cite{futurepaper}).  

\begin{figure}[htb]
\psfull
 \begin{center}
  \leavevmode
  \epsfig{file=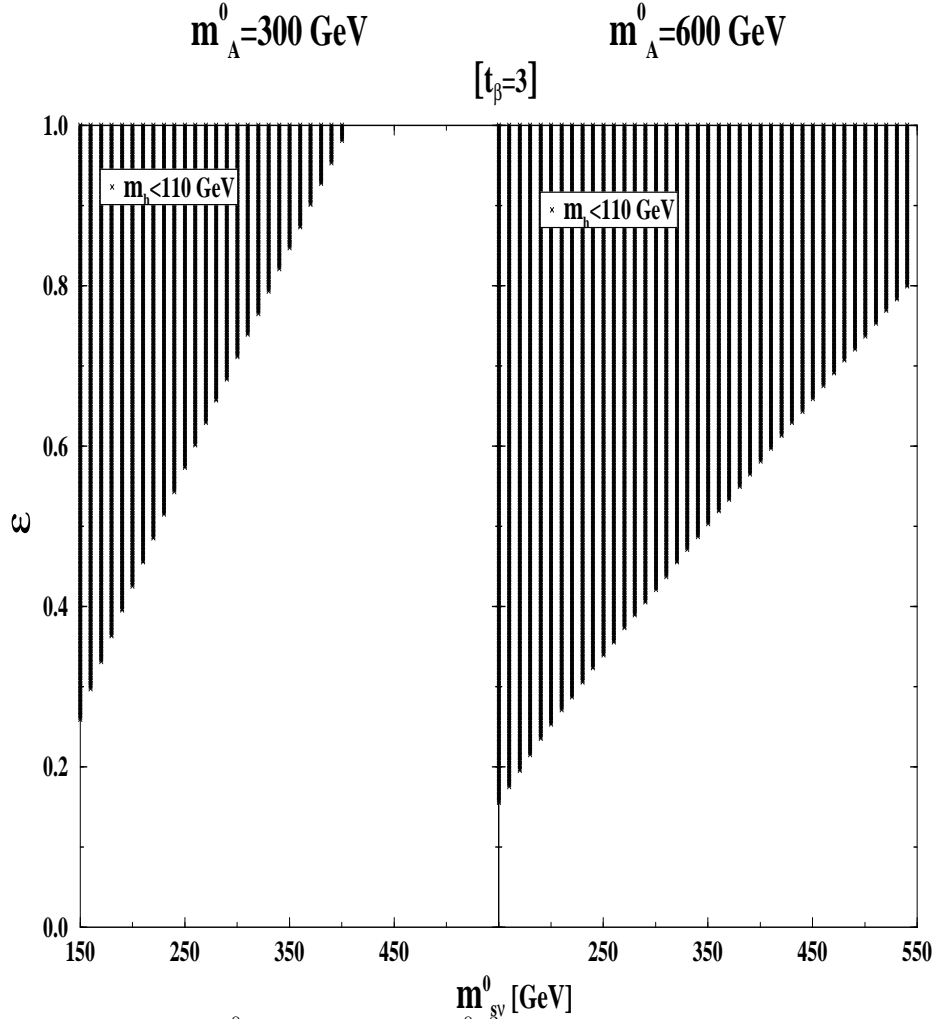,height=12cm,width=15cm,bbllx=0cm,bblly=2cm,bburx=20cm,bbury=25cm,angle=270}
 \end{center}
\caption{\emph{The shaded area in the $\varepsilon - m_{s \nu}^0$ plane
[$\varepsilon \equiv b_3 t_\beta/ (m_A^0)^2$] 
is excluded by the recent LEP2 limit on the light Higgs 
mass $m_h \gsim 110$ GeV. This excluded region 
is independent of $\lambda_{131}$.}}
\label{fig1}
\end{figure}

Moreover, the fact that the mass shifts due to $\varepsilon \neq 0$ 
are proportional 
to the sign of ($m_{\Phi_k^0} - m_{\Phi_\ell^0}$) has important consequences 
on the light CP-even 
Higgs particle. In particular,  
we find that if $m_A^0,m_{s \nu}^0 > m_{h^0}$ (as always chosen below), then 
$m_h$ tends to decrease with $\varepsilon$.
We can thus use the present LEP2 limit on $m_h$ 
to deduce the allowed range in e.g., the $\varepsilon - m_{s \nu}^0$ plane,  
for a given $m_A^0$. In particular, 
the present LEP2 bound is roughly 
$m_h \gsim 110$ GeV, for $m_A \gsim 200$ GeV 
irrespective of $t_\beta$  
and in 
the maximal mixing scenario with a typical SUSY scale/squark mass 
of 1 TeV \cite{lep2mh},\footnote{Since 
$b_3 \ne 0$ the $hZZ$ coupling [$A_2$ in (\ref{sigma})] is 
smaller than its value in the RPC case leading to a smaller 
$e^+e^- \to Zh$ production rate. The limits on $m_h$ given  
in \cite{lep2mh} are therefore slightly weaker in the RPV 
case (see e.g., \cite{davidson}).}  
Therefore, here after, we include the dominant higher order corrections 
(coming from the $t - \tilde t$ sector)
to the ($\xi_d^0,\xi_u^0$) block 
in $M_+^2$,
using the approximated formulae given in \cite{mhcorrections} 
with the maximal mixing scenario 
(as defined in \cite{mhcorrections}) 
and setting the typical squark mass at $m_{\tilde q} \sim 1$ TeV.
For example, in Fig.~\ref{fig1} we show the excluded region in the 
$\varepsilon - m_{s \nu}^0$ plane (the shaded area)
from the recent LEP2 limit of $m_h \gsim 110$ GeV which holds for 
the parameters set 
$\tan\beta=3$ and $m_A^0=300$ or $600$ GeV as used in Fig.~\ref{fig1}. 

In what follows, we focus on the case of a heavy Higgs 
spectrum, in particular $(m_A^0)^2 \gg M_Z^2$, which leads 
to the near mass degeneracy $m_H^0 \sim m_A^0$ 
and equivalently $m_H \sim m_A$.
We find that with $(m_A^0)^2 \gg M_Z^2$, 
the $\tilde\nu_+$ sneutrino-like state 
will potentially yield the dominant signal. 
This can be understood as follows: 
(i) A light Higgs ($h$) resonance in on-shell $VV$ pair production 
is theoretically excluded,
since the c.m. energy required to produce an on-shell 
$VV$ pair is at least $\sim 25$ GeV above    
the highest possible $m_h$
(the theoretical upper limit on $m_h$ is $\sim 135$ GeV).
Therefore, the $h$ contribution 
to $\sigma_V^0$ is always negligible and in particular 
near a $\tilde\nu_+$ 
resonance.\footnote{Note, however, that
in this scenario an $s$-channel $h$ exchange may 
lead to similar resonant enhancement in 
$e^+e^- \to V V^*$ at lower c.m. energies,  
where $V^*$ is an off-shell $W$ or $Z$.}
(ii) For $(m_A^0)^2 \gg M_Z^2$ and $\varepsilon \ll 1$, 
a heavy Higgs ($H$) resonance in $\sigma_V^0$ 
will also be much smaller than a $\tilde\nu_+$ resonance since

\begin{eqnarray}
S_{11} ~ \stackrel{\varepsilon \to 0}
{\longrightarrow} ~ \cos\alpha ~ \stackrel{m_A^2 \gg M_Z^2}
{\longrightarrow} ~ \sin\beta ~~~,~~~
S_{21} ~ \stackrel{\varepsilon \to 0}
{\longrightarrow} ~ \sin\alpha ~ \stackrel{m_A^2 \gg M_Z^2}
{\longrightarrow} ~ - \cos\beta ~,
\end{eqnarray}

\n leading to $A_1 \to 0$, where $A_1$ is the reduced $HVV$ coupling 
defined in (\ref{sigma}).
In addition, for $\varepsilon \to 0$ the element connecting $\tilde\nu_+^0$
to $H$ vanishes, i.e., $S_{31} \to 0$. 
Thus, since the $H$ exchange contribution 
to $\sigma_V^0$ is 
$\propto \Lambda_{H e^+e^-} \times \Lambda_{H V V} \sim  
S_{31} \times A_1$, it is doubly suppressed.  
 
The sneutrino-like state ($\tilde\nu_+$), on the other hand, 
has a much stronger (than $H$) coupling to 
the incoming electron since it couples 
to $e^+e^-$ through its dominant $\tilde\nu_+^0$ component. In particular, 
$S_{33} \to 1$ as $\varepsilon \to 0$.  
Thus, even though 
$\Lambda_{\tilde\nu_+ V V}$ and $\Lambda_{H V V}$ are comparable
(or equivalently $A_3 \sim A_1$),\footnote{Since 
$(M_+^2)_{13}/(M_+^2)_{23} = t_\beta$, $\tilde\nu_+^0$ acquires 
a larger $\xi_d^0$ mixing (than a $\xi_u^0$ mixing) 
which in turn implies a larger $H$ mixing, 
since the $H$ mass-eigenstate is mostly the $\xi_d$ weak-state 
when $(m_A^0)^2 \gg m_Z^2$ and $t_\beta^2 \gg 1$.}
the $\tilde\nu_+$ exchange contribution to $\sigma_V^0$, being 
$\propto S_{33} \times A_3$, will be more pronounced than the $H$ 
exchange one since $S_{33} \gg S_{31}$ for $\varepsilon \ll 1$. 

Therefore,  
the more favorite scenario for observing such a sneutrino-Higgs mixing 
resonance in $VV$ pair production is when the
$\tilde\nu_+$ resonates. 
It should be noted, however, 
that $\sigma_V^0$ may be further enhanced considerably
if both $m_{s \nu}$ and $m_H$ happen to lie close to the c.m. energy in 
the given experiment. As mentioned before, we do not consider in this paper 
such a possibility of an accidental mass degeneracy between the $H$ and 
$\tilde\nu_+$ states which may give rise to  
a ``combined'' $H + \tilde\nu_+$ resonance.      
Hence, in what follows we will consider only 
the case of a sneutrino-like resonance in $e^+e^- \to VV$.

The $\tilde\nu_+$ width, $\Gamma_{\tilde\nu_+}$ in (\ref{sigma}), needs to 
be included, since it controls the behavior of $\sigma_V^0$ in the vicinity
of our $\tilde\nu_+$ resonance. Assuming that the lightest neutralino 
($\tilde\chi_1^0$) is the 
Lightest SUSY Particle (LSP)
and also that $m_{s \nu} > m_{\tilde\chi_1^+}$, where 
$\tilde\chi_1^+$ is the lighter chargino, then the RPC two-body decays
$\tilde\nu_+ \to \tilde\chi_1^0 \nu_\tau,~ \tilde\chi_1^+ \tau$ are open 
and dominate. 
Indeed, for $m_A^2 \gg M_Z^2$ and 
following the traditional assumption 
of an underlying grand unification with a common gaugino 
mass parameter of $m_{1/2} < m_{s \nu}$, the mass hierarchy 
$m_{\tilde\chi_1^0} < m_{\tilde\chi_2^0} \sim 
m_{\tilde\chi_1^+} < m_{s \nu}$ and $m_{\tilde\chi_{3,4}^0} \sim 
m_{\tilde\chi_2^+} > m_{s \nu}$ is possible, e.g., when 
$m_{s \nu} < m_A$ \cite{djuadi}. Thus,
upon ignoring 
phase space factors, a viable 
conservative estimate 
is (see e.g., Barger {\it et al.} in \cite{previousrpv} 
and \cite{ourcpsnu}): 
$\Gamma_{\tilde\nu_+} \sim 
\Gamma (\tilde\nu_+ \to \tilde\chi_{1,2}^0 \nu_\tau) +     
\Gamma (\tilde\nu_+ \to \tilde\chi_1^+ \tau) \sim 10^{-2} 
m_{s \nu}$, which we use below. Note that 
for the ranges of $\varepsilon,~m_A^0$ 
and $m_{s \nu}^0$ considered 
the possible 
RPV decays are sufficiently smaller and  
$\Gamma_{\tilde\nu_+} \sim \Gamma_{\tilde\nu_+^0}$ since $S_{33} \to 1$.
Also, for reasons explained above, 
$\Gamma_H$ and $\Gamma_h$ have a negligible effect 
on the $\tilde\nu_+$ resonance and are therefore neglected.

Before presenting our numerical results we note the following:
(i)
Sufficiently away from threshold ($\beta_V \to 1$), 
$\sigma_W^0/\sigma_Z^0 \sim (\delta_W c_W^2/\delta_Z)\cdot (M_Z/M_W)^2 
\sim 2$ and, since typically $\sigma_W^{SM}/\sigma_Z^{SM} > 10$, 
the relative effect 
of the scalar exchange cross-section is more pronounced in 
the $ZZ$ channel. 
(ii) As mentioned above, for $\varepsilon \ll 1$ and in the decoupling 
limit [i.e., $(m_A^0)^2 \gg m_Z^2$], $\Lambda_{hVV} \to 1$ and 
$\Lambda_{HVV} \to 0$. At the same time, 
when $t_\beta^2 \gg 1$, $\xi_d^0 \to H$ and
$\xi_u^0 \to h$ so that, accordingly, for 
$t_\beta^2 \gg 1$, 
$\left(\Lambda_{\xi_u^0 VV}/\Lambda_{\xi_d^0 VV} \right) \gg 1$.
Thus, since $(M_+^2)_{23} = b_3 = \varepsilon (m_A^0)^2/t_\beta$,
the $\tilde\nu_+^0 - \xi_u^0$ mixing decreases with $\tan\beta$ 
and so 
as $t_\beta$ increases      
the sneutrino ``prefers'' to mix more with $\xi_d^0$ which 
has a suppressed coupling to $VV$ in this limit.
As a consequence, the sneutrino-like resonance effect in 
$\sigma_V^0$ drops with $\tan\beta$ in the limit of small RPV and 
$m_A^2 \gg M_Z^2$.   

\begin{figure}[htb]
\psfull
 \begin{center}
  \leavevmode
  \epsfig{file=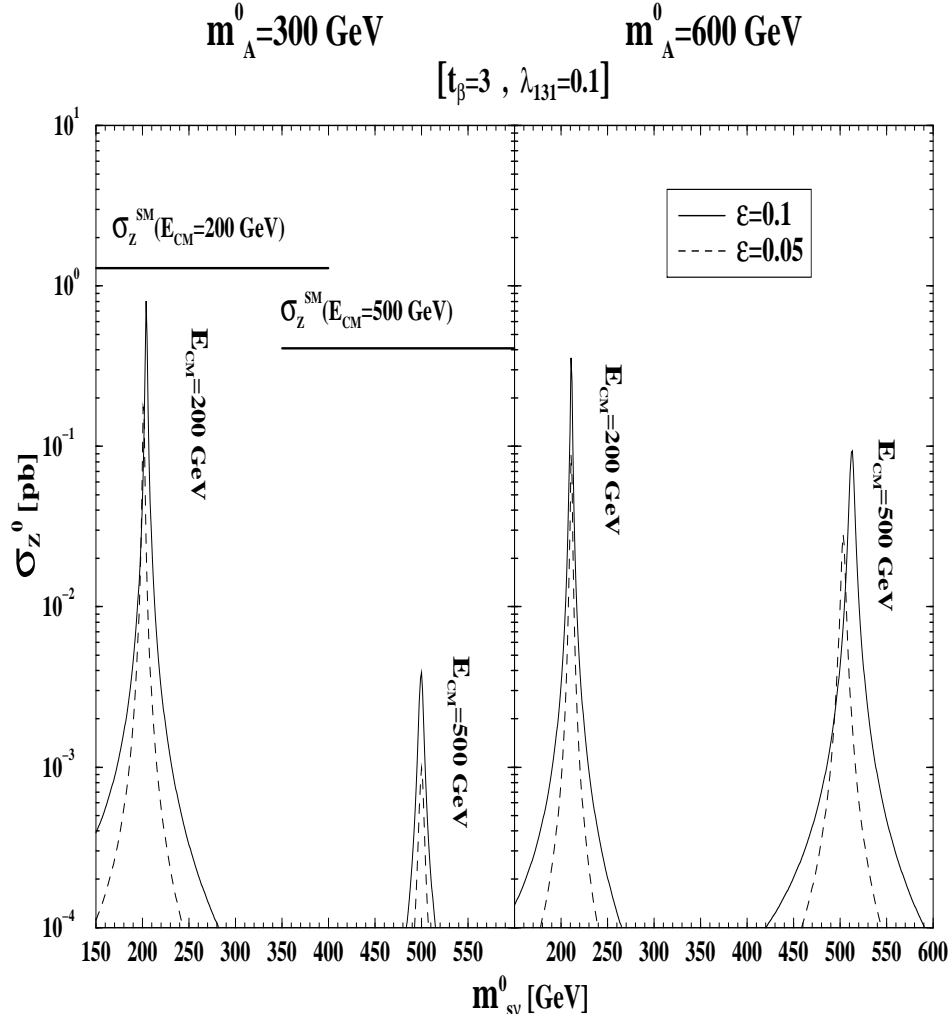,height=12cm,width=15cm,bbllx=0cm,bblly=2cm,bburx=20cm,bbury=25cm,angle=270}
 \end{center}
\caption{\emph{$\sigma_Z^0$ as a function of 
$m_{s \nu}^0$, for $m_A^0=300$ GeV (left plot) 
and $m_A^0=600$ GeV (right plot).
For both values of $m_A^0$, $\sigma_Z^0$ is shown for 
the c.m. energies $\sqrt s=200$ GeV with $\varepsilon=0.1,~0.05$ (left curves) 
and $\sqrt s=500$ GeV with $\varepsilon=0.1,~0.05$ (right curves).
$\lambda_{131}=0.1$ is used 
(note that $\sigma_Z^0$ scales as $\lambda_{131}^2$). 
The SM $ZZ$ cross-sections for $\sqrt s=200$ and $500$ GeV are also shown 
by the horizontal solid lines.}}
\label{fig2}
\end{figure}

In Fig.~\ref{fig2} we show
$\sigma_Z^0$ as a function of $m_{s \nu}^0$ for 
c.m. energies of $\sqrt s=200$ and 500 GeV.
This is shown for $t_\beta=3$ and for $m_A^0=300$ GeV (left side) or 
$m_A^0=600$ GeV (right side) 
(a more detailed investigation of the parameter space involved
will be given in \cite{futurepaper}). 
For definiteness we take $\varepsilon=0.05,~0.1$ and 
$\lambda_{131}=0.1$.\footnote{$\sigma_V^0$ is insensitive to 
the signs of $\varepsilon$ and $\lambda_{131}$.} 
The SM cross-sections 
$\sigma_Z^{SM}(\sqrt s=200~{\rm GeV}) \sim 1.29$ [pb] and 
$\sigma_Z^{SM}(\sqrt s=500~{\rm GeV}) \sim 0.41$ [pb]
are also shown by the horizontal solid lines. 

We see that, as expected, $\sigma_Z^0$ is larger for 
a smaller $|m_A^0 - m_{s \nu}^0|$
mass splitting, since the sneutrino--Higgs mixing phenomena is 
proportional to factors of $[(m_A^0)^2 - (m_{s \nu}^0)^2]^{-1}$
(see discussion above).  
Clearly, the scalar exchange cross-section can be
statistically significant even if the mass of the sneutrino-like 
scalar is away from the resonance, i.e., within a range of  
$m_{s \nu} - \sqrt s \leq \Delta$, where, as we shell see below, 
$\Delta$ may range from a few GeV to a few tens of GeV depending 
on $\varepsilon$ and the rest of the SUSY parameter space involved.  

\begin{figure}[htb]
\psfull
 \begin{center}
  \leavevmode
  \epsfig{file=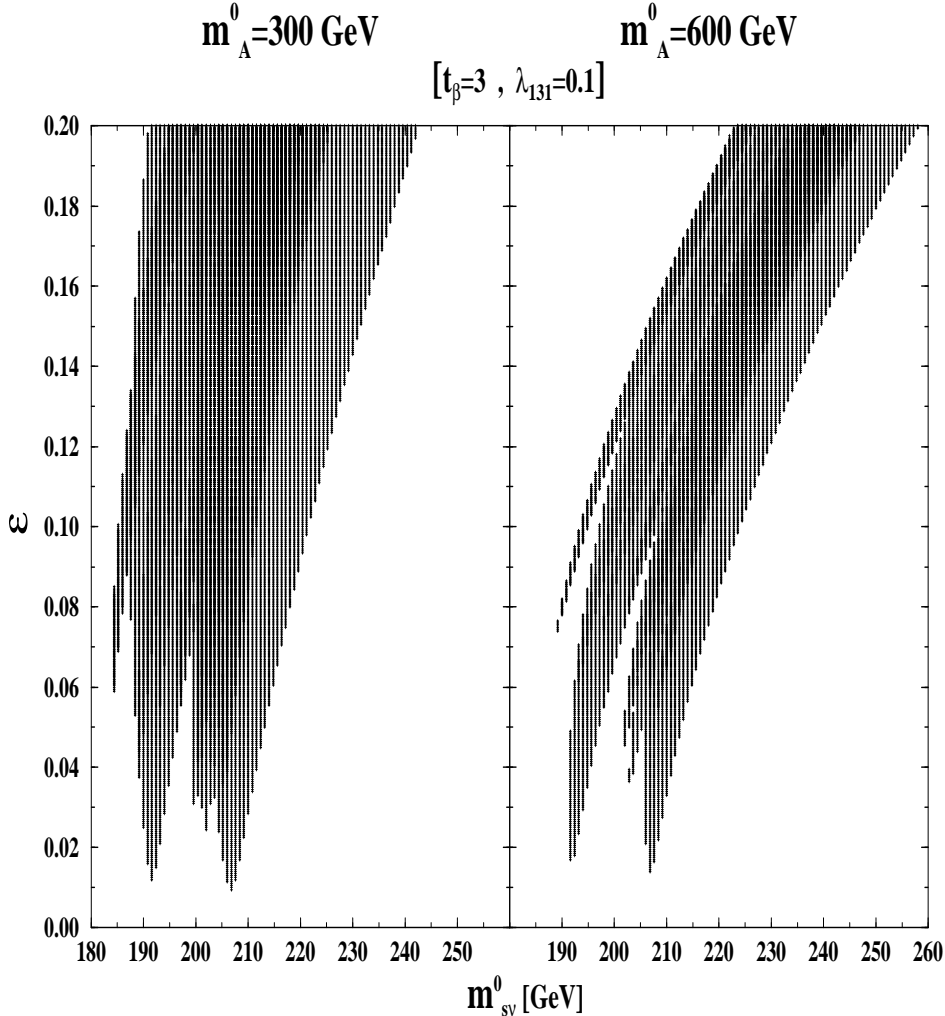,height=12cm,width=15cm,bbllx=0cm,bblly=2cm,bburx=20cm,bbury=25cm,angle=270}
 \end{center}
\caption{\emph{$1 \sigma$ excluded regions in the 
$\varepsilon - m_{s \nu}^0$ plane from the 
LEP2 measurements of the $WW$ and $ZZ$ cross-sections (see text).}}
\label{fig3}
\end{figure}

Thus, for the case of $\sqrt s$ around 200 GeV, we can 
use the measured values of the $WW$ and $ZZ$ 
cross-sections at LEP2 to place further bounds on the $\varepsilon - m_{s \nu}^0$ 
plane for a given $m_A^0$ and $t_\beta$. This is shown in Fig.~\ref{fig3}
where we have again set $m_A^0=300$ GeV or $m_A^0=600$ GeV, 
$t_\beta=3$, $\lambda_{131}=0.1$ and 
used 
the measured $\sigma_Z$ and $\sigma_W$, combined by the 4 LEP 
experiments, 
from the 
183, 189, 192, 196, 200, 202, 205 and 207 GeV LEP2 runs as given in 
\cite{lep2sigs}.
In particular, for each run we take the measured and the 
SM cross-sections (also given in \cite{lep2sigs}),\footnote{For 
the $ZZ$ and $WW$ SM cross-sections 
we use the results of the 
ZZTO and YFSWW3 Monte-Carlos, 
respectively, where we take 
a $2\%$ theoretical error for the ZZTO prediction and no error for 
the YFSWW3 one, see \cite{lep2sigs}.}
$\sigma_V^{exp;SM} \pm \Delta \sigma_V^{exp;SM}$, and require
that $\sigma_V^0 < (\sigma_V^{exp} - \sigma_V^{SM}) + 
\sqrt{(\Delta \sigma_V^{exp})^2+(\Delta \sigma_V^{SM})^2}$.\footnote{We do 
not include the cases in which 
$(\sigma_V^{exp} - \sigma_V^{SM}) + 
\sqrt{(\Delta \sigma_V^{exp})^2+(\Delta \sigma_V^{SM})^2} <0$.}

Evidently, the limits coming from 
the $ZZ$ and $WW$ cross-sections measurements give further restrictions  
at low $\varepsilon$ values below $\sim 0.2$ 
(in a sneutrino mass range of several tens of GeV)\footnote{Note that, 
since $b_3 = \varepsilon 
(m_A^0)^2/t_\beta$, 
these $1 \sigma$ limits
can be directly translated into limits on the $b_3 - m_{s \nu}^0$ plane.},
for which 
there are no bounds coming from the LEP2 limits on $m_h$ 
(see Fig.~\ref{fig1}). Note that the fingers like shape of the 
shaded area in Fig.~\ref{fig3} is an artifact of the fact that we are using 
a discrete set of c.m. energies in accordance with the LEP2 runs. 

Alternatively, for the case of a 500 GeV $e^+e^-$ collider 
we can find the mass range of the sneutrino-like scalar 
for which its contribution to the $WW$ and $ZZ$ cross-sections
 may be observable with a statistical 
significance of at least $3 \sigma$ by requiring 
$(\sigma_V^0 \sqrt{{\cal L}}/\sqrt{\sigma_V^0 + \sigma_V^{SM}})>3$. 
For example, we find that with an integrated 
luminosity of ${\cal L}=100$ fb$^{-1}$,  
$m_A^0=600$ GeV and $t_\beta=3$, 
a more than $3\sigma$ signal 
can arise in the $ZZ$ case 
within the sneutrino mass ranges
$490~{\rm GeV} \lsim m_{s \nu} \lsim 509~{\rm GeV}$ and 
$495~{\rm GeV} \lsim m_{s \nu} \lsim 505~{\rm GeV}$ for
$\varepsilon \sim 0.1$ and $0.05$, respectively. 
The corresponding $3 \sigma$ 
mass intervals in the $WW$ case are  
typically a factor $\sim 1.5$ smaller for the same values of
$t_\beta,~m_A^0$ and $\varepsilon$.  
These $3 \sigma$ mass ranges are further enlarged if an angular cut 
on the c.m. scattering angle, $\theta$, 
is imposed. For example, with $0 \lsim \cos\theta \lsim 1$, 
we find that, for $m_A^0=600$ GeV, $t_\beta=3$ and  
$\varepsilon \sim 0.1$ or $0.05$, 
the sneutrino resonance will be observable 
at $3 \sigma$ in the $WW$
channel within the mass ranges 
$489~{\rm GeV} \lsim m_{s \nu} \lsim 511~{\rm GeV}$ or 
$495~{\rm GeV} \lsim m_{s \nu} \lsim 505~{\rm GeV}$, respectively. 
These mass ranges are comparable to the ones obtained in the $ZZ$ case
with no angular cut.  

To summarize, a small lepton number violation scenario, 
which incorporates small trilinear and bilinear RPV terms 
into the SUSY Lagrangian, 
can lead 
to a significant scalar resonance enhancement in 
$e^+e^- \to ZZ,~W^+W^-$, due to mixings between the sneutrino and the Higgs 
particles, which 
may be accessible to a 500 GeV $e^+e^-$ collider.   
We also find that useful limits can be 
placed on this scenario from the LEP2 measurements of the 
$WW$ and $ZZ$ cross-sections and from the LEP2 limits on the 
light Higgs mass.  
Finally, we note that a similar scalar 
resonance may arise in top-quark
pair production due to the sneutrino--Higgs mixing phenomena (see 
\cite{futurepaper}). Such a resonance enhancement in 
$e^+e^- \to t \bar t$ should give further evidence in favor of 
the bilinear RPV SUSY scenario since the absence of a tree-level 
sneutrino--top--anti-top trilinear RPV coupling and the fact that 
the Higgs-electron-positron coupling is $\propto m_e$, makes the 
sneutrino--Higgs mixing the only viable mechanism for generating 
an observable resonance signal in $e^+e^- \to t \bar t$ within the 
SUSY framework.

\bigskip
\bigskip

We thank D. Guetta for discussions. 
G.E. thanks the
U.S.-Israel Binational Science Foundation, the Israel Science
Foundation and the Fund for Promotion of Research at the
Technion for partial support.

\end{document}